\PassOptionsToPackage{hyphens}{url}
\PassOptionsToPackage{unicode=true}{hyperref}
\documentclass[a4paper,11pt]{article}
\pdfoutput=1

\usepackage{jcappub}
\usepackage[T1]{fontenc} 
\usepackage{xcolor}
\usepackage{booktabs}
\usepackage{multirow}
\usepackage{textalpha}
\usepackage{siunitx}
\usepackage{orcidlink}
\usepackage{bm}
\DeclareUnicodeCharacter{2009}{\,}

\newcommand{\lya}{Ly$\alpha$}
\newcommand{\poned}{$P_{\mathrm{1D}}$}

\title{\boldmath Generation of Lognormal Synthetic Lyman-$\alpha$ Forest Spectra for $P_{1D}$ Analysis}

\author[a]{Meagan Herbold\,\orcidlink{0009-0000-8112-765X}}
\author[a]{Naim G{\"o}ksel Kara{\c c}ayl{\i}\, \orcidlink{0000-0001-7336-8912}}
\author[a]{Paul Martini\, \orcidlink{0000-0002-4279-4182}}

\affiliation[a]{\textit{Department of Physics, The Ohio State University, \\
191 West Woodruff Ave, Columbus, OH 43210, U.S.A.}}

\emailAdd{herbold.9@osu.edu}
\emailAdd{karacayli.1@osu.edu}
\emailAdd{martini.10@osu.edu}

\abstract{
The one-dimensional flux power spectrum (\poned) of the Lyman-$\alpha$ forest probes small-scale structure in the intergalactic medium (IGM) and is therefore sensitive to a variety of cosmological and astrophysical parameters. These include the amplitude and shape of the matter power spectrum, the thermal history of the IGM, the sum of neutrino masses, and potential small-scale fluctuations due to the nature of dark matter. However, \poned\ is also highly sensitive to observational and instrumental systematics, making accurate synthetic spectra essential for validating analyses and quantifying these effects, especially in high-volume surveys like the Dark Energy Spectroscopic Instrument (DESI). We present an efficient lognormal mock framework for generating one-dimensional Lyman-$\alpha$ forest spectra tailored for \poned\ analysis. Our method captures the redshift evolution of the mean transmitted flux and the scale-dependent shape and amplitude of the one-dimensional flux power spectrum by tuning Gaussian field correlations and transformation parameters. Across the DESI Early Data Release (EDR) redshift range ($2.0 \leq z \leq 3.8$), and a wide range of scales ($10^{-4}$ s km$^{-1} \leq k \leq 1.0$ s km$^{-1}$), our mocks recover the mean flux evolution with redshift to sub-percent accuracy, and the \poned\ at the percent level. Additionally, we discuss potential extensions of this framework, such as the incorporation of astrophysical contaminants, continuum uncertainties, and instrumental effects. Such improvements would expand its utility in ongoing and upcoming surveys and enable a broader range of validation efforts and systematics studies for \poned\ inference and precision cosmology.

\vskip 1em
\noindent\textbf{Keywords:} Lyman alpha forest, power spectrum, quasars: absorption lines

\vskip 0.5em
\noindent\textbf{ArXiv ePrint: } \href{https://arxiv.org/abs/2509.04405v1}{2509.04405}

\vskip 0.5em
\noindent\textbf{DOI: } \href{https://doi.org/10.1088/1475-7516/2026/01/031}{10.1088/1475-7516/2026/01/031}
}

\begin{document}
\maketitle
\flushbottom

\section{Introduction}

    The Lyman-$\alpha$ (\lya) forest is a distinctive series of absorption features seen in the spectra of distant galaxies and quasars, detectable from ground-based instruments over redshifts approximately spanning the end of reionization ($z \sim 6$) to the peak of galaxy formation ($z \sim 2$) \cite{McQuinn2016}. As light from background sources travels through regions of moderate gas overdensity, such as those associated with foreground galaxies and the intergalactic medium (IGM), neutral hydrogen absorbs photons at the \lya\ resonance and this process imprints redshifted absorption lines on the spectra of background objects \cite{GunnPeterson1965, Weinberg2013}. These characteristic absorption lines trace fluctuations in the matter density field and provide multiple independent samples of redshift and distance along the line of sight. This makes the forest a powerful probe of density fluctuations and large-scale structure.

    Early work in \lya\ forest cosmology demonstrated the forest’s sensitivity to the underlying matter distribution, prompted the study of the matter power spectrum, and the subsequent investigation and constraints on several small-scale effects such as power suppression due to warm dark matter and neutrinos \cite{Croft_1998, McDonald_2006, Seljak2005}. The Baryon Oscillation Spectroscopic Survey (BOSS) \cite{Dawson2013}, a part of the Sloan Digital Sky Survey (SDSS), obtained the first detection of baryon acoustic oscillations (BAO) in the \lya\ forest region at high redshift ($z \sim 2$) and enabled a new tool for cosmological distance measurements at these redshifts \cite{Busca_2013, Slosar_2013}. Numerous studies continued to analyze larger samples through the end of the extended Baryon Oscillation Spectroscopic Survey (eBOSS) \cite{Dawson2016}, which culminated with significant improvements in \lya\ results  \cite{du_Mas_des_Bourboux_2020}, and now continues with state-of-the-art dark energy surveys such as the Dark Energy Spectroscopic Instrument (DESI) \cite{DESI2016a.Science, DESI2016b.Instr}. DESI has already substantially improved the statistical precision of cosmological measurements, and continues to explore the nature of dark energy with the largest map of the known universe.  DESI has already obtained the most precise measurement of the expansion history to date with \lya\ \cite{DESI-DR1-LyA,DESI-DR2-I} and lower-redshift galaxies and quasars \cite{DESI-2024-VI,DESI-DR2-II}. 

    The one-dimensional flux power spectrum (\poned) of the \lya\ forest is particularly sensitive to a wide range of astrophysical parameters and processes, especially at the small scales that are typically inaccessible to traditional galaxy survey techniques such as weak lensing and clustering \cite{Karacayli_2024, Viel2006}. This sensitivity enables the power spectrum to probe and constrain a variety of interesting phenomena, such as the primordial power spectrum \cite{Viel2004, Seljak2005}, neutrinos \cite{PD15_constraints, PD15_mass_and_Cosmo, PD20}, dark matter \cite{Baur_2016, Ravoux2023, Irsic2017, Viel2013, Villasenor2023}, and the thermal properties of the IGM \cite{Walther_2018, Boera2019, Villasenor2022}. As measurements of the \lya\ forest improve in size and precision, the use of synthetic spectra for thorough analysis validation is increasingly necessary for robust cosmological inference.  

    A critical component of any cosmological measurement is the thorough and accurate characterization of systematic errors, such as instrumental effects and astrophysical contaminants. This is especially necessary for \poned\ analysis, where a variety of systematics such as high column density (HCD) systems, damped \lya\ absorbers (DLA), continuum fitting uncertainties, metal-line contamination, as well as noise and resolution, can significantly reduce the constraining power of power spectrum measurements \cite{Karacayli_2020}.
    
    Hydrodynamical simulations are necessary to investigate how varying initial conditions and theoretical models evolve over time to form cosmic structures and systems \cite{Weinberg1997,Viel2006,Almgren_2013, Rogers2017,Villasenor2021}. Such simulations can provide detailed, high-fidelity realizations of the IGM by modeling gas dynamics, thermal feedback, and nonlinear gravitational collapse. However, simulations are computationally expensive, that is they require substantial time, memory, and processing power. Therefore, it is often impractical to use hydrodynamical simulations to generate the large ensembles of spectra needed for robust error estimation, pipeline validation, and the exploration of systematic effects. N-body simulations, while less costly, still require additional post-processing to model \lya\ absorption. And emulator-based approaches, which interpolate between a set of high-resolution measurements from hydrodynamical simulations, can offer improved speed but may be limited by the coverage and granularity of the training grid \cite{Walther_2018,Bird_2019,Cabayol_2023,Walther_2025}.

    Lognormal spectral simulations offer a fast and analytically tractable alternative for generating the large volumes of semi-realistic synthetic spectra or mocks that are needed for analysis validation. The lognormal approximation provides a computationally efficient means to model the non-Gaussian, nonlinear distribution of neutral hydrogen column densities, capturing essential features of the \lya\ forest while remaining several orders of magnitude faster than other approaches. While not as detailed or physically motivated as hydrodynamic simulations--and therefore insufficient for full cosmological inference--lognormal methods are nonetheless well suited for validation testing, where speed and control are critical. The utility of the lognormal approach to generating mock spectra is particularly relevant for current and future stages of DESI. Lognormal mocks have already been employed for \lya\ \poned\ analysis of the DESI Early Data Release (EDR) \cite{Ravoux2023,Karacayli_2024} and Data Release 1 (DR1) \cite{Karacayli_2025,Ravoux_2025}. Yet the greater statistical precision in upcoming releases will require even more accurate and flexible synthetic datasets. Lognormal mocks remain a powerful tool in this context, offering a practical balance between computational speed, control over target observables, and the ability to quantify systematic effects \cite{Karacayli_2020}.    

    Lognormal mocks have played a crucial supporting role in modeling the one- and three-dimensional correlation functions of the \lya\ forest in large spectroscopic surveys, such as BOSS \cite{Font-Ribera_2012, Bautista_2015}, eBOSS \cite{Chabanier_2019,Farr_2020,Etourneau_2024}, and now DESI \cite{Karacayli_2020, Herrera_Alcantar_2025, casas2025},  ensuring that the precision gained by larger volumes of data is not limited by systematics rather than statistical uncertainty.
    Beyond the \lya\ forest, lognormal methods have found growing application in a variety of cosmological contexts. They have been employed to investigate properties of the IGM like thermal history and Jeans length \cite{ondro2024}, to emulate weak lensing shear and convergence fields for pipeline validation \cite{zhong2024}, and to efficiently model matter fields and large-scale structure \cite{Xavier2016, Tessore_2023}. These broader applications underscore the versatility of the lognormal formalism, particularly in regimes where rapid generation of controlled realizations is essential for assessing systematic effects or testing analysis methods at scale.
    
    This study builds on the lognormal mock framework first presented in McDonald et al. (2006) \cite{McDonald_2006}, and further developed by Kara{\c{c}}ayl{\i} et al.\ (2020) \cite{Karacayli_2020}. We have adapted and extended that method for efficiently generating synthetic \lya\ forest spectra to better replicate key observational properties--specifically, the redshift evolution of the mean transmitted flux and the one-dimensional flux power spectrum. By introducing tunable Gaussian field correlations and modifying transformation parameters, we improve the fidelity of the mock spectra and enable finer control over target observables. We demonstrate that our improved lognormal method substantially reduces discrepancies in \poned\ reconstruction of target models or datasets compared to previous-generation mocks and enables reliable modeling for future DESI datasets. 
        
    The outline of the paper is as follows. In section \S \ref{section: method} we describe the general method used to generate lognormal synthetic spectra, as well as how we fit the \poned\ and mean flux to match observational datasets. We then present the results and limitations of the mock measurements, as well as a comparison with observational data, in \S \ref{section: Results}. We discuss potential applications and extensions of our method in \S \ref{section: Discussion}, and summarize our conclusions in \S \ref{Section: conclusion}. Additional information on software and data availability is provided in Appendices \ref{section: software} and \ref{Section: Data Availability}, respectively, and further details about the \poned\ fitting method and assumptions necessary for the calculation are in Appendix \ref{section: covariance matrix and variance}.

\section{Method} \label{section: method}
    The aim of this work is to produce synthetic \lya\ forest spectra that closely reproduce key observational statistics, including the mean transmitted flux and one-dimensional flux power spectrum. These realistic mocks allow for thorough testing and validation of analysis pipelines, helping to ensure that cosmological inferences drawn from real data are not biased by methodological or systematic effects. This section presents an improved methodology for generating such mocks, building upon the lognormal framework established in previous work \cite{McDonald_2006, Karacayli_2020, Karacayli_2024}. Earlier methods relied on a fixed analytic form for the power spectrum of the underlying Gaussian field, often held constant across redshift and tuned empirically to approximately match the observed \poned. While effective at limited redshifts and scales, and specific to particular datasets, such static forms are limited in precision and flexibility. This is especially crucial in the current era of precision cosmology, where measurements extend to higher redshifts and smaller scales.

    The key innovation in our method is a new approach to modeling \poned\ that does not assume a fixed input power shape. Instead, we solve directly for the optimal Gaussian correlation function $\xi_G$ in configuration space such that the resulting flux correlation function $\xi_F$--after applying the full sequence of lognormal transformations--reproduces the target flux \poned. We perform this inversion point-wise by comparing the transformed flux correlation function to a smoothed version of the target model and identifying the excess correlation (or power) needed in the Gaussian field to recover the desired flux statistics. Once we determine the best-fit $\xi_G$, we Fourier transform it to produce a redshift-dependent Gaussian power spectrum $P_G(k,z)$, which replaces the static analytic formulation used in prior work. This method provides a more accurate and flexible framework for generating mocks that reproduce the detailed scale and redshift dependence of the \lya\ forest power spectrum.

    The remainder of this section is structured as follows. We begin in Section \ref{section: Mock_generation} by describing the pipeline for generating lognormal mock spectra, which closely follows the method of Kara{\c c}ayl{\i} et al. (2020) \cite{Karacayli_2020}. We then present our new method for solving the Gaussian input power spectrum that drives the shape of \poned\ and evaluate its performance relative to previous methods in Section \ref{Section: fitting_power}. Finally, we detail the procedure to fit the evolution of the mean transmitted flux with redshift in Section \ref{Section: fitting_mean_flux}.

\subsection{Lognormal Mock Generation} \label{section: Mock_generation}

    Our baseline modeling approach aims to match the \poned\ measured by Kara{\c c}ayl{\i} et al. (2023) \cite{Karacayli_2024} and the optical depth measurement of Turner et al. (2024) \cite{turner24}, both of which use data from DESI and are discussed further in \S \ref{Section: fitting_power} and \S \ref{Section: fitting_mean_flux}, respectively. 
    
\begin{figure}[ht]
       \centering
       \includegraphics[width=0.7\linewidth]{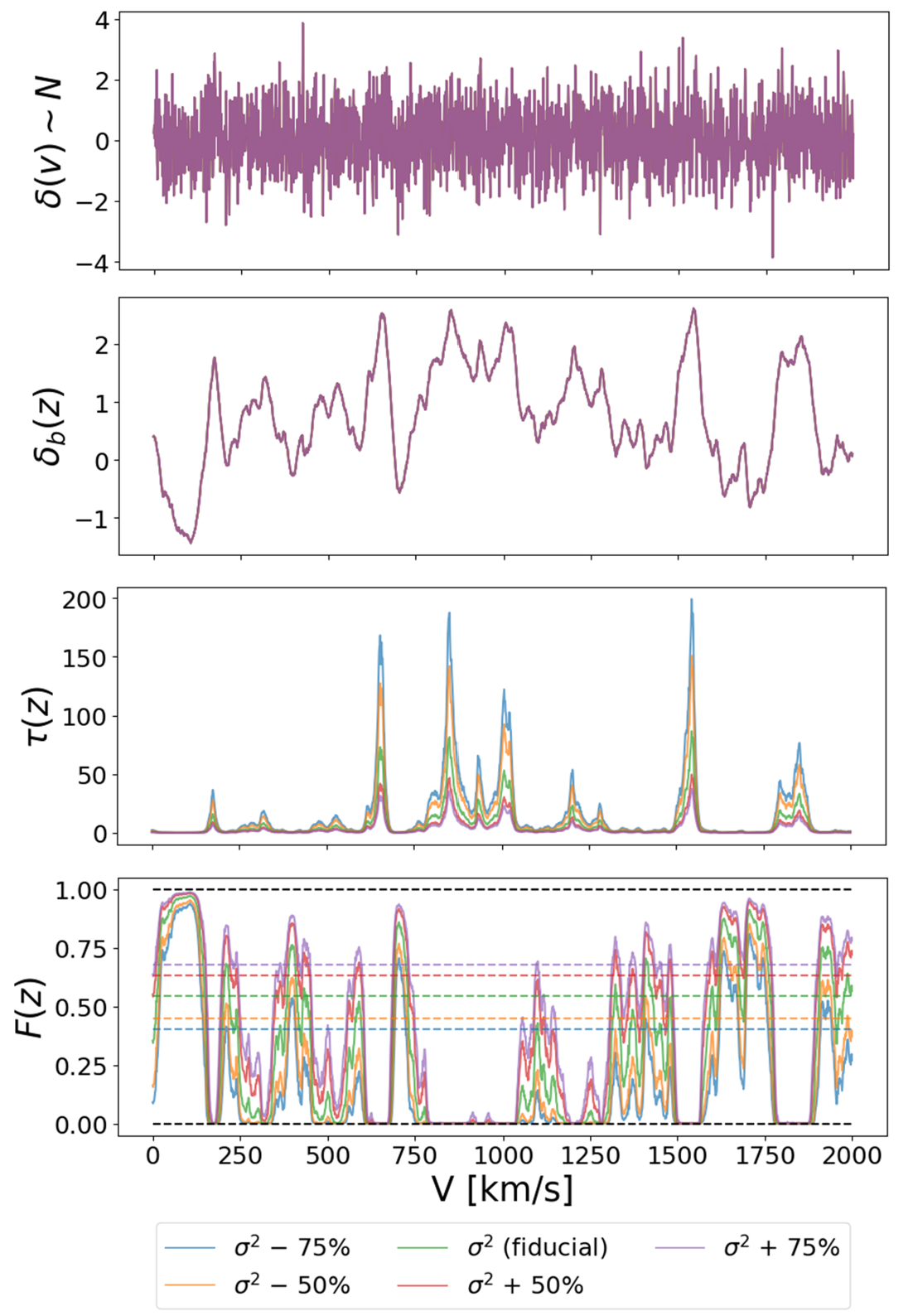}
       \caption{An example line-of-sight realization at redshift z = 3.6, illustrating the impact of varying an individual parameter value during mock generation. The value of the variance, $\sigma_F^2$, is shown spanning a $\pm75\%$ range around the best-fit (fiducial) value given in Table \ref{table.2}. Each row, from top to bottom, shows a key step in the transformation pipeline: the Gaussian random field, the redshifted baryon fluctuation approximation $\delta_b(z)$, the optical depth $\tau(z)$, and the final transmitted flux $F(z)$.}
       \label{fig.1}
   \end{figure}

    Figure \ref{fig.1} illustrates a representative line-of-sight mock spectrum at key stages of the generation pipeline, highlighting the effects of modifying individual input parameters. We begin by generating a one-dimensional random Gaussian velocity grid, which is shown in the first panel of Figure \ref{fig.1}. Key features of this grid include even spacing, normal distribution (zero mean and unit variance), and sufficiently high resolution. We then perform a Fourier transform on the grid and apply a baseline model power spectrum by multiplying the transformed grid by $\sqrt{P_G(k)/dv}$, where $dv$ is the spacing of the initial grid in velocity space, and $P_G(k)$ is the power spectrum associated with the underlying Gaussian field. 

    In contrast to previous work, which relied on a single analytic form for $P_G(k)$ that held parameters fixed across all redshifts, our method computes a unique Gaussian power spectrum for each redshift. We achieve this by solving for a corresponding Gaussian correlation function, $\xi_G$, which directly shapes the amplitude and scale-dependence of the final \poned. Further details of this redshift-dependent fitting procedure for $P_G(k,z)$ are provided in \S\ref{Section: fitting_power}.
    
    Finally, we apply an inverse Fourier transform to obtain $\delta_b(v)$, which is an approximation of the underlying baryon fluctuations. The variance of this field $\sigma_b^2(v)$ characterizes the strength of baryonic fluctuations in the delta field. 
    
    We define a redshift evolution factor $a(z)$ by: 
        \begin{equation}
            a^2(z) = \left( \frac{1+z}{1+z_0} \right) ^{-\nu}, \label{eq.2}
        \end{equation}
    where $z_0$ is the pivot redshift and $\nu$ is one of the tuning parameters used to fit the mean flux. The pivot redshift $z_0 = 3.0$, and the pivot mode $k_0 = 0.009$ km s$^{-1}$, are adopted from the BOSS power spectrum measurements by \cite{PD13}. A redshift evolved delta field can then be obtained by multiplying $\delta_b(v,z) = a(z)\delta_b(v)$, which is shown in the second panel of Figure \ref{fig.1}. The variance of the baryonic fluctuations $\sigma_b^2(v)$, and the flux fitting parameter $\sigma_F^2$, the value of which is determined in \S \ref{Section: fitting_mean_flux}, are redshifted separately by multiplying by $a^2(z)$.

    To approximate the non-linearity and non-Gaussianity of the {\sc HI} column density field, we apply a squared lognormal transformation: 
        \begin{equation}
            n(z) = e^{2 \delta_b(z)-\sigma_F^2(z)}. \label{eq.3}
        \end{equation} 
     Note that the variance used in the lognormal transformation is the free parameter $\sigma_F^2(z)$, determined by fitting the mean flux, rather than that of the delta field, $\sigma^2_b(z)$, which varies both with redshift and individual spectrum realizations. We also define an optical depth evolution factor $t(z)$:
        \begin{equation}
            t(z) = \tau_0 \left( \frac{1+z}{1+z_0} \right) ^{\tau_1},  \label{eq.4}
        \end{equation}
    where $\tau_0$ and $\tau_1$ are additional mean flux fitting parameters. The grid is then transformed to optical depth by multiplying $\tau(z) = t(z)  n(z)$. This is shown in the third panel of Figure \ref{fig.1}. Finally, the optical depth is transformed to the transmitted flux by $F(z) = e^{-\tau(z)}$, which is shown in the final panel of Figure \ref{fig.1}.
        
    The mean flux and power spectrum can be derived analytically using the one-point and two-point probability, respectively. The mean flux $\overline{F}(z)$ can be calculated with the following expression, which has been formulated for efficient evaluation using Gauss-Hermite quadrature integration:
        \begin{equation}
            \overline{F}(z) = \frac{1}{\sigma_b \sqrt{2\pi}}\int_{-\infty}^{\infty}\exp \left[ -\frac{\delta^2}{2\sigma_b^2} - x(z) \text{e}^{2a(z)\delta} \right] \text{d}\delta, \label{eq.8}
        \end{equation}
    where $\delta$ is the base Gaussian random field and $e^{-\delta^2/2\sigma^2}$ is the probability density of the field. Here, $x(z)$ is defined as: 
        \begin{equation}
            x(z) \equiv \tau_0 \left( \frac{1+z}{1+z_0} \right)^{\tau_1} \text{e}^{-a^2(z)\sigma_F^2}.\label{eq.9}
        \end{equation}
        
    To analytically calculate the power spectrum, we first define fluctuations of the flux field as $\delta_F(v,z) = F(v,z)/\overline{F}(z) -1$. The correlation function of the flux fluctuations is then defined as $\xi_F = \langle \delta_F(x) \delta_F(x+r) \rangle$, which can be rewritten as $\xi_F  = \langle F_iF_j/\overline{F}^2 \rangle-1$. The correlation function at a given redshift can then be expressed as an integral over two independent Gaussian $\delta$ fields as 
    \begin{equation}
        1 + \xi_F(v_{ij}) = \int \frac{\text{e}^{- \bm{\delta}^T \textbf{C}^{-1} \bm{\delta} / 2}}{2 \pi \sqrt{\text{det}\textbf{C}}} \frac{F_i F_j}{\overline{F}^2} \text{d}\bm{\delta}, \label{eq.10}
    \end{equation}
    where
    \begin{equation}
        \textbf{C} =    \begin{pmatrix}
                        \sigma^2 & \xi_G(v_{ij}) \\
                        \xi_G(v_{ij}) & \sigma^2 
                        \end{pmatrix} \label{eq.11}
    \end{equation}
    is the covariance matrix and $\xi_G(v)$ is the correlation function of the underlying Gaussian field. The relation between this correlation function $\xi_G$ and the flux correlation function $\xi_F$ is crucial to fitting the properties of the one-dimensional power spectrum, as discussed in \S \ref{Section: fitting_power}. We recover the \poned($k$) by taking the inverse Fourier transform of the flux correlation function  $\xi_F$ and normalize by the spacing of the velocity grid. 

    Although the framework described above produces general lognormal mock spectra, additional steps are required to ensure that the simulations match observational data. In particular, the mocks must be tuned to reproduce the mean transmitted flux and the shape of the \poned. These properties serve as critical benchmarks for mock validation and for assessing the fidelity of analysis pipelines when applied to survey data.

   An example implementation of this mock generation method is available in the GitHub repository referenced in Appendix~\ref{section: software}.

\subsection{One-Dimensional Power} \label{Section: fitting_power}

    The generation of mocks that match the observed \poned\ requires some modification to the approach presented in the previous subsection. We achieve this by solving for the optimal Gaussian correlation function $\xi_G$ that--after undergoing the series of nonlinear transformations described in \S \ref{section: Mock_generation}--produces a flux correlation function $\xi_F$ that is consistent with the target \poned. Rather than relying on a fixed analytic form for the input power, we numerically solve for a discrete set of correlation values, $\xi_G(v)$, at each redshift. These are then transformed into Gaussian power spectra, $P_G(k,z)$, providing a unique and redshift-specific input power spectrum tailored to match the target flux statistics. 
    
    We start from a smooth, high-resolution representation of the target flux power spectrum, transform it into configuration space to obtain $\xi_{F,\mathrm{target}}$, and iteratively adjust the input $\xi_G$ until the transformed $\xi_F$ matches the target. This procedure effectively solves for the excess correlation (or equivalently, excess power) that must be introduced in the Gaussian field in order to reproduce the desired flux statistics. Once we obtain the best-fit $\xi_G$, we Fourier transform it to determine the redshift-dependent Gaussian power spectrum $P_G(k,z)$ that serves as the baseline input to the mock generation pipeline. This data-driven, redshift-specific inversion significantly improves agreement with the target power spectrum across a wide range of scales and redshifts.
    
    The default power model, which is motivated by the formalism in \cite{McDonald_2006,PD13}, is:
        \begin{equation}
            \frac{k P(k,z)}{\pi} = A \frac{(k/k_0)^{3+n+\alpha ln k/k_0}}{1+(k/k_1)^2}\left( \frac{1+z}{1+z_0} \right)^{B + \beta ln k/k_0}, \label{eq.7}
        \end{equation}
    where $k_0 = 0.009$ km/s$^{-1}$ is the pivot wavenumber and $z_0 = 3.0$ is the pivot redshift \cite{PD13,Baur_2016,Karacayli_2020}, where the pivot point correction ($k_0,$) is motivated by \cite{Gontcho_2014}. This baseline power estimate model is based on the method used by \cite{PD13}, and was later modified by \cite{Karacayli_2020} with the addition of a Lorentzian decay. When this model is applied to the DESI EDR \lya\ forest data \cite{RamirezPerez_2024}, the best-fit parameters for eq. \ref{eq.7} are: $A = 0.076$, $n = -2.521$, $\alpha = -0.13$, $B = 3.67$, $\beta = 0.29$, and $k_1 = 0.037$ \cite{Karacayli_2024}. 
    
    We first interpolate the target \poned\ model to achieve uniform spacing and high resolution on a one-dimensional grid. The smoothed target is then inverse Fourier transformed and normalized by the velocity spacing of the grid to obtain the target correlation function, $\xi_{F,target}$.  We then sample $\xi_{F,target}$ logarithmically in velocity to enhance computational efficiency and provide a smoother fit in log space.

    To solve for $\xi_{F,target}$, we utilize the relationship between the flux correlation function $\xi_F$ and the underlying Gaussian field $\xi_G$, which is defined in eq.\ \ref{eq.10} and eq.\ \ref{eq.11}, along with the transformations detailed in \S \ref{section: Mock_generation}. We determine the optimal values of the Gaussian correlation function $\xi_{G,fit}$, whose transform produces a flux correlation function $\xi_{F,fit}$ that closely resembles $\xi_{F, target}$. This optimization is performed point-wise for each $\xi_{F, target}$ value using \textsc{scipy.optimize.least\_squares}\cite{2020SciPy-NMeth}. The constrained non-linear solver iteratively adjusts $\xi_{G,fit}$ to minimize the residual between $\xi_{F,fit}$ and the corresponding $\xi_{F,target}$.
    
    We interpolate the optimal $\xi_{G,fit}$ to ensure uniform spacing and high-resolution sampling and then calculate another Fourier transform to obtain a Gaussian power spectrum, $P_G(k,z)$. This power spectrum represents the baseline power associated with the underlying Gaussian field, which is used during the mock generation process in \S \ref{section: Mock_generation}.

\begin{figure}[ht]
        \centering
        \includegraphics[width=1.0\linewidth]{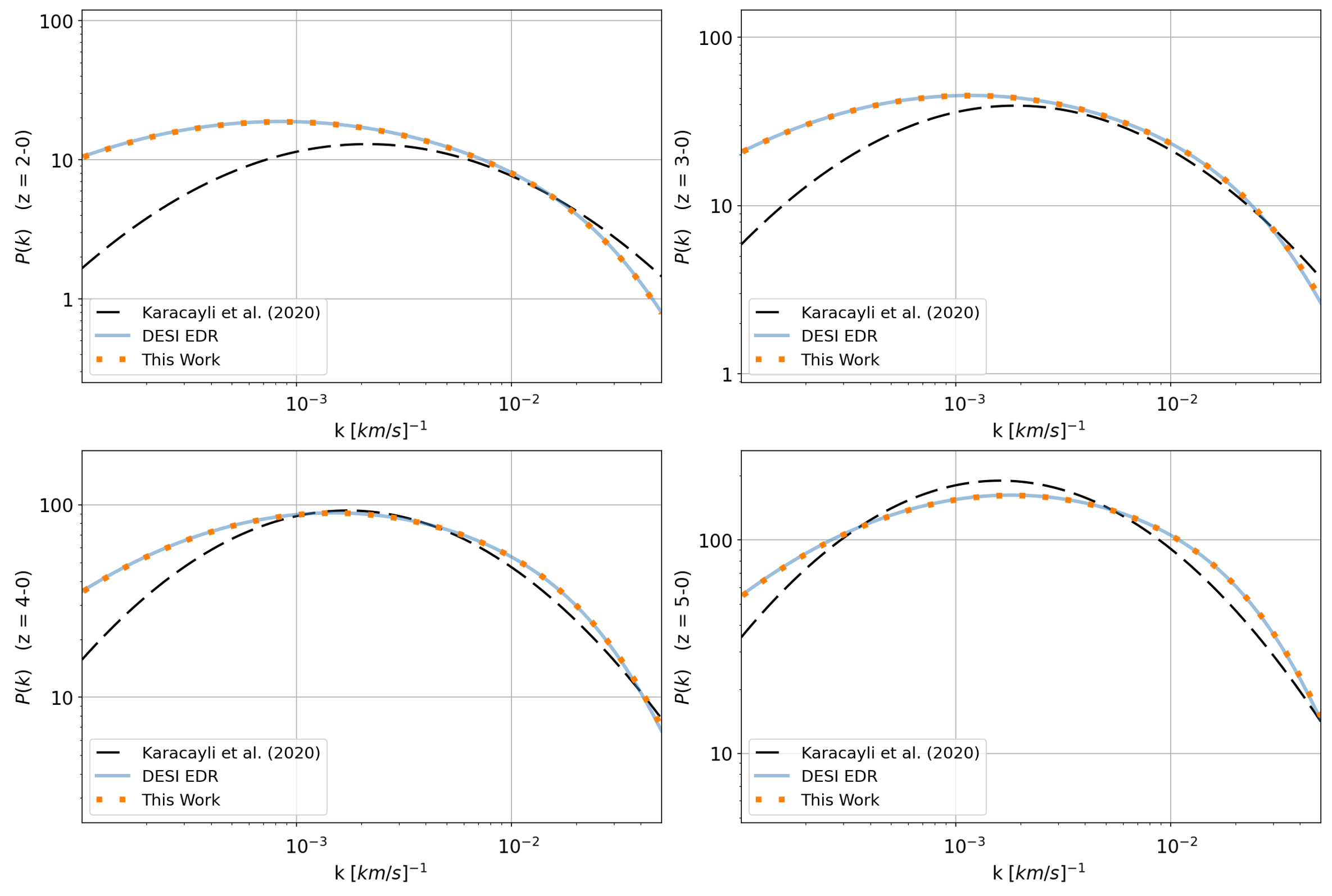}
        \caption{Comparison of the \poned\ of mocks generated using the method presented in this work with the DESI EDR measurement and the previous mock generation method. DESI EDR (blue, solid line) is a fit of eq. \ref{eq.7} to the \poned\ measurement by Kara{\c{c}}ayl{\i} et al. (2024) \cite{Karacayli_2024}. This Work (orange, dotted) shows the \poned\ produced with our method, which agrees with the DESI EDR fit within 1\%.  
        The previous method by Kara{\c{c}}ayl{\i} et al.\ (2020) \cite{Karacayli_2020} (black, dashed) is also shown. The four panels are redshifts $z = 2, 3, 4,$ and $5$. } \label{fig.3}
    \end{figure}

    The performance of our approach is illustrated in Figure \ref{fig.3}. Here we used the best fit correlation function $\xi_{F,\text{fit}}$, derived from our analytical solution, to calculate the \poned\ at representative redshifts in the range $2.0 \leq z \leq 5.0$. Each panel also shows the best fit \poned\ power to the DESI EDR measurement (used as the target for our method) and the previous mock generation method \cite{Karacayli_2020}. The predicted \poned\ from our method closely tracks the shape and amplitude of the DESI EDR target model on all scales. The recovered power remains within 1\% of the input at intermediate redshifts for all scales, and there are only slightly larger deviations at the highest and lowest redshifts due to the covariance matrix approximation (see Appendix \ref{section: covariance matrix and variance}).

\subsection{Mean Flux} \label{Section: fitting_mean_flux}

    Modeling the evolution of the mean transmitted flux of the \lya\ forest is essential for characterizing the statistical properties of the forest and has been the subject of extensive study. The mean flux can provide a critical benchmark for validating mock spectra, as it directly reflects the cumulative absorption along the line of sight and underpins many cosmological and astrophysical interpretations. While early measurements relied on composite quasar spectra and statistical continuum fitting techniques \cite{Faucher-Giguere_2008, Becker2013}, recent advances have utilized large spectroscopic datasets and machine learning methods, such as convolutional neural networks (CNNs), to improve continuum predictions and reduce systematic uncertainties across a broad range of redshifts \cite{turner24}.

    Although presented after the \poned\ section, we fit the mean flux first because it sets the overall amplitude and clustering strength of the lognormal flux field. We determine these global absorption properties with a small number of tunable parameters that are analytically accessible via one-point statistics, which makes them significantly more straightforward to constrain than the full shape of the power spectrum. Once we match the mean flux, the remaining small-scale and scale-dependent discrepancies in the power spectrum (i.e., the residual structure not captured by the mean flux alone) are addressed through the method described in \S \ref{Section: fitting_power}. This two-step approach separates the modeling of the average opacity from that of the detailed flux correlations, while ensuring consistency between both sets of observables.

    To ensure that our mocks reproduce the observed mean flux evolution, we adopt the effective optical depth model from Turner et al. (2024) \cite{turner24}, derived from DESI DR1. This model expresses the mean optical depth as a smooth redshift-dependent power law, $\tau(z) = \tau_0 (1+z)^\gamma$, with the corresponding flux given by $\overline{F}(z) = e^{-\tau(z)}$. We then fit four key parameters, $\tau_0$, $\tau_1$, $\nu$, and $\sigma_F^2$, within the mock generation framework described in \S \ref{section: Mock_generation} to match this target mean flux model. Here, $\tau_0$ sets the amplitude of the optical depth and corresponds to $\tau(z=z_0)$, where $z_0$ is the pivot redshift. The redshift scaling is governed by $\tau_1$, which controls how the optical depth evolves relative to $z_0$, while $\nu$ determines the rate of the redshift evolution factor $a^2(z)$. Finally, $\sigma_F^2$ sets the variance of the base Gaussian field, $\delta(v)$.
    
    We optimize these parameters by minimizing a weighted least-squares cost function, ensuring that the simulated flux field reproduces the amplitude and redshift evolution of the observed mean absorption to sub-percent accuracy over the redshift range $2\lesssim z \lesssim 5$.

    We define a chi-squared-like cost function for the fit:  
        \begin{equation*}
            \chi^2 = \sum_i  \frac{\overline{F}_{\text{model}}(z_i) - \overline{F}_{\text{fit}}(z_i)}{\sigma_{\text{model}}^2},
        \end{equation*}
    where $cost = d \cdot d$, $\overline{F}_\text{model}(z_i)$ is the desired mean flux model as a function of redshift ($z$), such as the default DESI-like model described above \cite{turner24}, and $\overline{F}_{\text{fit}}(z_i)$ is the mean flux measured by eq. \ref{eq.8} for a single line-of-sight realization generated with a given set of values for the fitting parameters $\tau_0$, $\tau_1$, $\nu$, and $\sigma_F^2$. The uncertainty term, $\sigma^2_{\text{model}} = 0.1\%  \ \ \overline{F}_{\text{model}} +1\text{e}^-5$, which accounts for both the relative precision of the model flux and a small constant floor to ensure numerical stability.
    
   We determine the optimal values for these fitting parameters with a local minimizer, \textsc{iminuit}\cite{iminuit}, which are then used for mock generation as detailed in \S \ref{section: Mock_generation}. These values, along with their associated Hesse errors, are listed in Table \ref{table.2}. The Hesse errors provide a measure of uncertainty for each parameter and are computed from the inverse of the Hessian matrix of the objective function, evaluated near the minimum. 

   \begin{table}[ht]
       \centering
       \begin{tabular}{|c | c c|}
            \hline
             & Value & Hesse Error \\
            \hline
            $\tau_0$   &  0.67377  &  6.0e-05 \\
            $\tau_1$   &  5.31008  &  1.6e-4  \\
            $\nu$      &  2.16175  &  1.7e-4  \\
            $\sigma_F^2$&  1.50381  &  1.4e-4  \\
            \hline
       \end{tabular}
       \caption{Optimal parameter values and associated Hesse errors for the generation of mock spectra with DESI-like mean flux \cite{turner24}, following the method detailed in \S \ref{section: Mock_generation}.}
       \label{table.2}
   \end{table}

    The effects of varying an individual fitting parameter for an example line-of-sight realization are shown in Figure \ref{fig.1}. The figure demonstrates the effect of altering the value of the variance, $\sigma_F^2$, at various stages of the mock generation progress. The values of the fiducial parameters correspond to the optimal values for generating spectra that match the mean flux evolution observed with DESI, given in Table \ref{table.2}. 
    
    The most significant changes in the transmitted flux fraction occur for variations in $\nu$ and $\tau_1$, particularly at redshifts far from the pivot redshift $z_0 = 3.0$. This behavior reflects the strong dependence on redshift introduced by equations \ref{eq.2} and \ref{eq.4}. In contrast, variations in $\tau_0$ and $\sigma_F^2$ primarily affect the overall amplitude of the final flux transmission field, with relatively less sensitivity to the pivot redshift. These parameters directly influence the normalization of the optical depth and the strength of density fluctuations, respectively, leading to changes in absorption strength.
   
   \begin{figure}[ht]
       \centering
       \includegraphics[width=0.9\linewidth]{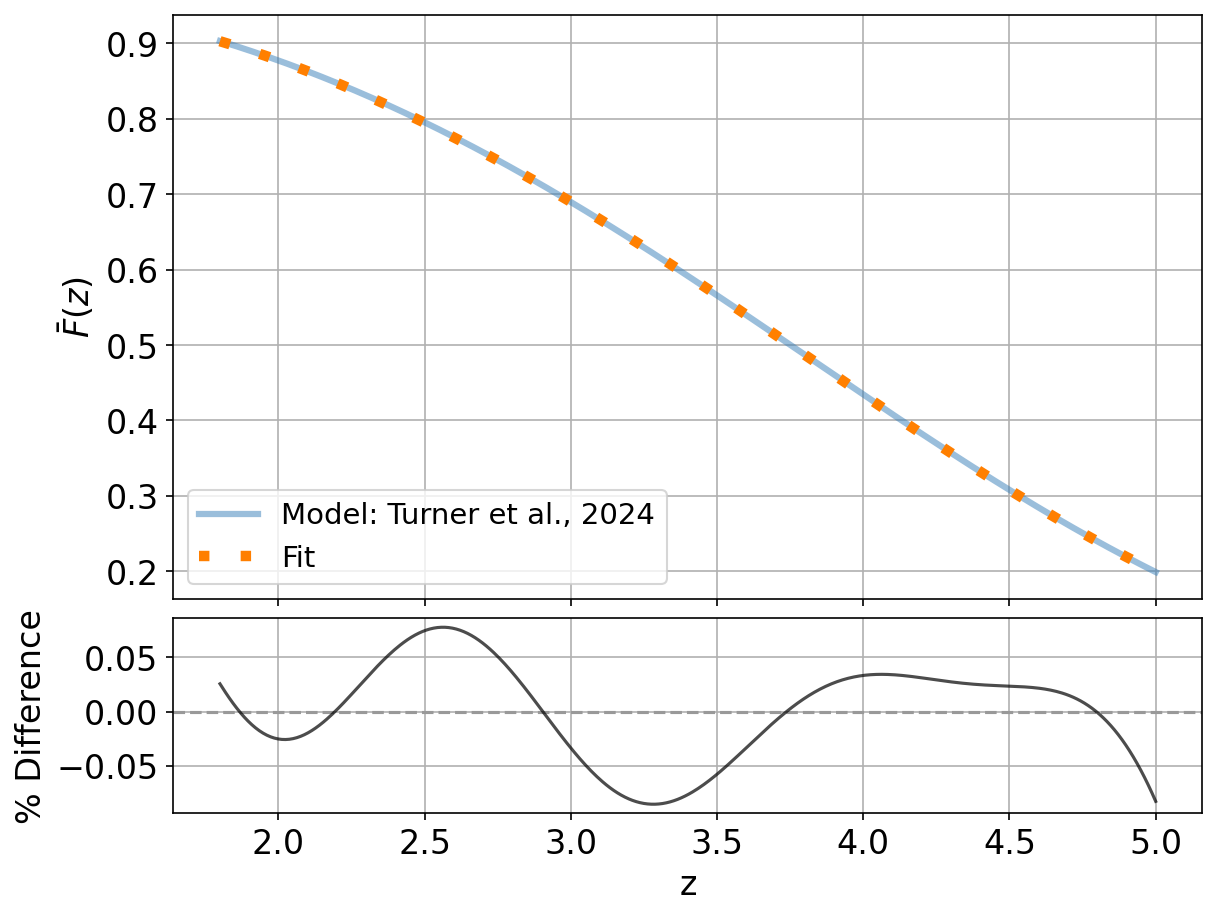}
       \caption{Comparison of the analytic model for the mean flux from mocks generated in this work to a model fit to observational data.  (Top) Comparison between the default mean flux redshift evolution model fit by Turner et al. (2024)\cite{turner24} to DESI DR1 (blue) and the best-fit analytic model of this work, using the derived parameter values in Table \ref{table.2} (orange, dotted) over the redshift range $2 \lesssim z \lesssim 5$. (Bottom) Percent difference between the DESI mean flux measurement and the our analytic prediction. 
       }
       \label{fig.2}
   \end{figure}   

    In contrast to the physically motivated procedure of Ondro et al.(2025) \cite{ondro2024}, our procedure cannot be used for the inference of the thermal history of the IGM. While the Ondro et al. method relies on an external photoionization rate measurement to derive the mean flux of the IGM, our lognormal transformations do not aim to model the underlying physics from first principles, but rather to match observables through tuning of four transformation parameters. Whether these four parameters can then be used to constrain the equation of state of the IGM, we leave to future work.
        
\section{Results}\label{section: Results}

    We evaluate the accuracy of our fitting procedure by generating 1000 independent line-of-sight realizations of synthetic spectra at distinct redshifts spanning the same range as the DESI EDR \poned\ measurement ( $2.0 \leq z \leq 3.8$ ) \cite{Karacayli_2024}. The choice of 1000 realizations is motivated by a convergence analysis of the RMS fractional error, described in Appendix~\ref{section: Convergence Analysis}. From these uncontaminated transmitted flux spectra, we compute the mean transmitted flux and \poned\ using Equations \ref{eq.8} and \ref{eq.10}, respectively. 
    
\begin{figure}[ht]
        \centering
        \includegraphics[width=1.0\linewidth]{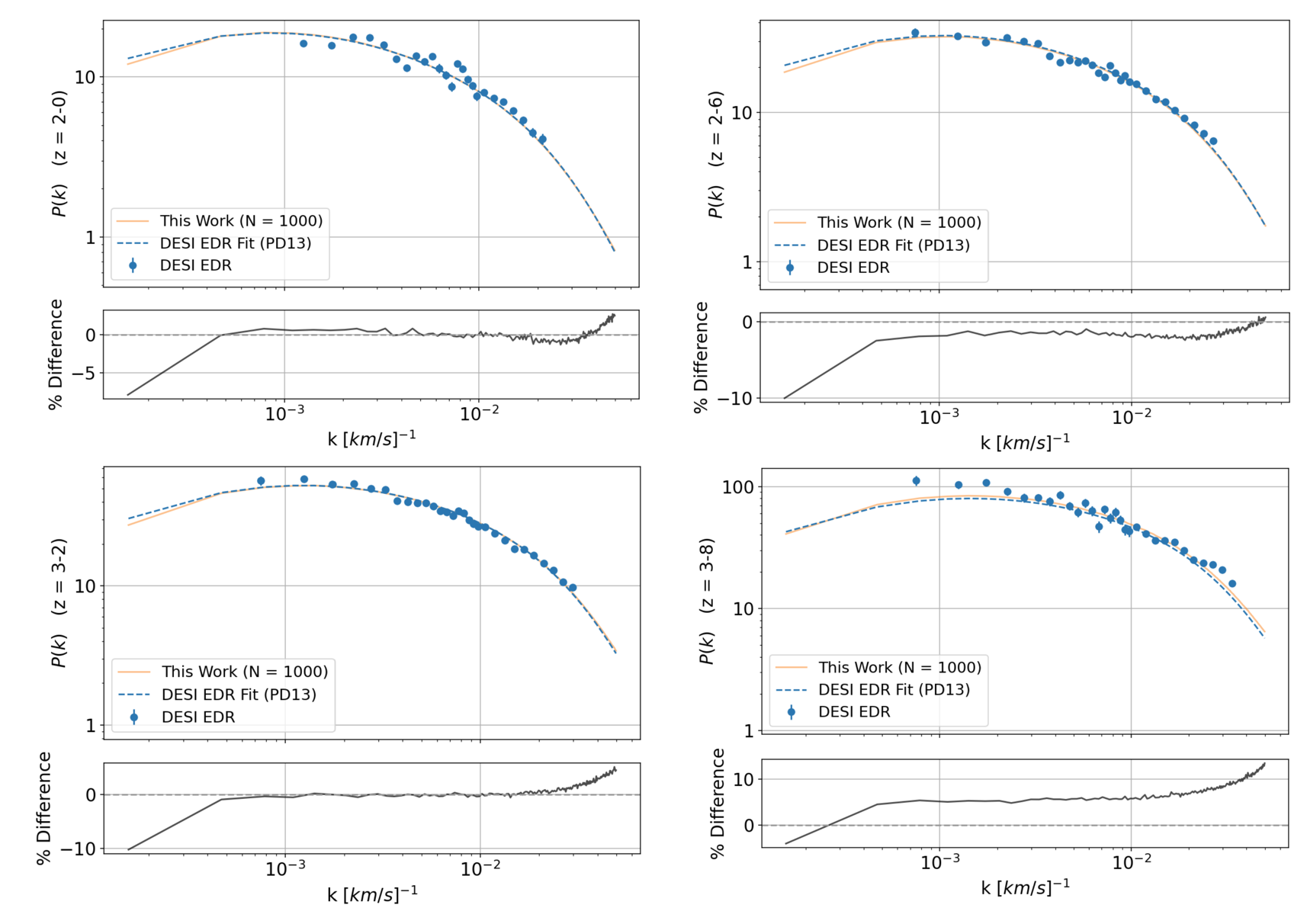}
        \caption{A demonstration of how well mocks generated using our method recover the input power spectrum target model. (Top) Comparison of the default \poned\ model, given by eq. \ref{eq.7}  fit to the DESI EDR \poned\ measurement \cite{Karacayli_2024} (blue, dashed), with the binned \poned\ measurement averaged over 1000 line-of-sight realizations per redshift (orange). This comparison is done for select redshifts $z = 2.0, 2.6$, $3.2$, and $3.8$, which spans the range of the DESI EDR measurement. (Bottom) Percent difference between the average measured power from our mocks and the input target model.
        }
        \label{fig:mock_power_measurement}
    \end{figure}

    Our mocks accurately recover both the shape and amplitude of \poned\ across the full DESI EDR redshift range, and over a broad range of scales  ($10^{-4} \, \mathrm{s\,km}^{-1} \leq k \leq 0.1 \, \mathrm{s\,km}^{-1}$) that extends beyond the $k$-range covered by the DESI measurements. This performance is illustrated in Figure~\ref{fig:mock_power_measurement}, which shows the average measured \poned\ from the mocks, compared to the best-fit of eq. \ref{eq.7} to the DESI EDR \poned measurements by Kara{\c c}ayl{\i}\ et al. (2024) and high-resolution \poned by Walther et al. (2018) \cite{Walther_2018}, which was used as the target to tune our mocks. Each panel presents results at a representative redshift, with 1000 independent line-of-sight realizations per redshift. In addition, the lower sub-panels display the percent difference between our mock measurements and the DESI EDR best-fit model. The residuals remain below 15\% across the full $k$-range considered, and below 10\% across the DESI EDR $k$-range, with somewhat larger discrepancies at the upper and lower redshift edges ($z \sim 2.0$ and $z \sim 3.8$), where our method slightly underestimates and overestimates the power, respectively. We attribute this small bias, which is especially prevalent at the smallest scales (large $k$) to simplifying assumptions in the covariance matrix that were used during the model-fitting procedure (see Appendix \ref{section: covariance matrix and variance}).

    To contextualize these improvements, we compare the root-mean-square (RMS)  fractional error between our method and the previous method at representative redshifts. This comparison highlights both the accuracy and flexibility with which our method can recover a target model or dataset, regardless of scale. We begin by evaluating performance over the $k$-ranges used in DESI-Lite \cite{Karacayli_2020} and DESI EDR \cite{Karacayli_2024}, followed by a comparison using the redshift-dependent $k$-limits defined in the DESI DR1 measurement \cite{Karacayli_2025}. Finally, we test our model over an even wider $k$-range to assess its scalability for future datasets.

    Over the DESI-Lite $k$-range ($0.0005 \leq k \leq 0.112$ s km$^{-1}$), our mocks yield an average RMS fractional error of 0.16, an improvement over the 1.12 error from the previous method. Similarly, when limiting the range to reflect conservative bounds that match the lowest and highest bins from the DESI EDR measurement ($0.00075 \leq k \lesssim 0.035$ s km$^{-1}$), our method achieves an RMS error of 0.02, compared to 0.10 from the earlier approach. While the absolute improvements appear modest, the gains are especially significant at the smallest and largest scales, as shown in Figure \ref{fig.3}. For example, at $z = 2.0$, our maximum deviation is just 1.3\%, whereas the earlier method deviates by up to 44.8\% at low $k$ ($\sim 10^{-3}$ s km$^{-1}$).  

    The DESI DR1 \poned\ measurement adopts a dynamic $k$-range that adjusts with redshift, which aims to optimize information use at each redshift. Although $k_{min}$ is fixed due to continuum error contamination, the upper limit is defined by $k_{max}=\frac{0.5\pi}{R_z}$, where $R_z = \frac{c \Delta \lambda}{(1+z) \lambda_{Ly \alpha}}$, and $\Delta \lambda_{\text{DESI}}=0.8$ \AA, based on an estimate of the size of the spectrograph resolution correction. Using these redshift-dependent limits, we find an average RMS fractional error of 0.01 with our method, compared to 0.09 from the previous approach -- an improvement by nearly an order of magnitude. As before, the greatest improvements occur at the smallest $k$-values. At $z = 2.0$, our method deviates by a maximum of 1.3\%, while the earlier method reaches up to 37\% over the same range of scales.

    Finally, we evaluate performance over an extended $k$-range ($10^{-5} \leq k \lesssim 0.1$ s km$^{-1}$), testing the robustness of our model at the extremes of scale. Across the full redshift interval, our mocks maintain an average RMS fractional error of 0.11, a dramatic improvement over the 18.15 error when using the earlier method. Similar to our comparisons with the DESI measurement ranges, we find that the differences are once again most pronounced at the largest and smallest scales where the previous method exhibits limited fidelity in reproducing the shape of the power spectrum. Even in this wide range, the worst-case deviation of our method remains below 40\%, compared to 300\% for the previous approach, highlighting its scalability and suitability for high-precision datasets in future surveys.
    
\begin{figure}[ht]
        \centering
        \includegraphics[width=0.9\linewidth]{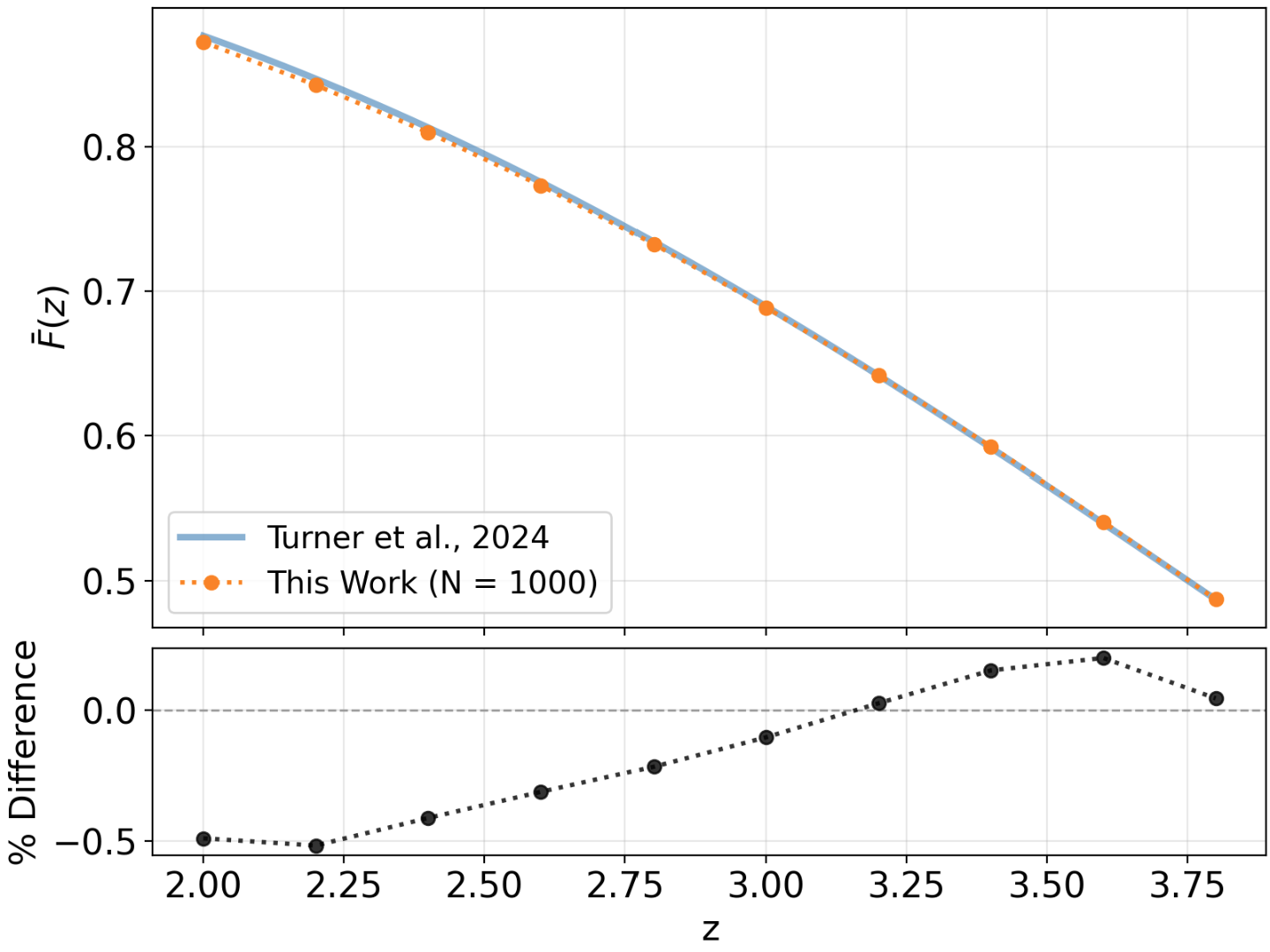}
   \caption{A demonstration of how well our mocks recover the input  mean flux model.  (Top) Average mean flux measured from a set of 1000 independent line-of-sight mock realizations at each target redshift in the range $2.0 \leq z \leq 3.8$ (orange, dotted) compared to the measurement by Turner et al., (2024) \cite{turner24} (blue) on DESI DR1. This DESI measurement was the target model used to tune the fitting parameters for mock generation in \S \ref{Section: fitting_mean_flux}. (Bottom) Percent difference between the DESI DR1 mean flux measurement \cite{turner24} and the average mean flux measured from our mocks.}
        \label{fig:flux_measurement}
    \end{figure}

    In addition to matching the \poned, an accurate reproduction of the mean transmitted flux is valuable for validating mock spectra intended for \lya\ forest analyses. We therefore evaluate how well our mocks replicate the mean flux evolution with redshift with the same set of mocks used to measure the \poned. We find that our mocks reproduce the redshift evolution of the mean flux measured by Turner et al. (2024) \cite{turner24}. This agreement is illustrated in Figure \ref{fig:flux_measurement}. The top panel shows the DESI DR1 mean flux measurements by Turner et al. (2024) and our mock measured values. The lower panel displays the percent difference between the DESI EDR measurements used as our input target model and our mock measurements. We find that our method produces an RMS fractional error in the mean flux of just 0.002 across the DESI EDR redshift range ($2.0 \leq z \leq 3.8$). The largest deviations occur at the limits of the redshift range, with a maximum percentage deviation of $0.5\%$.  

\section{Discussion} \label{section: Discussion}
    In \S\ref{section: Results}, we demonstrated substantial improvements in the accuracy of our mock spectra relative to the previous \poned\ mock generation method \cite{Karacayli_2020}. The accuracy in recovering the mean flux evolution with redshift arises from optimizing transformation parameters discussed in \S \ref{Section: fitting_mean_flux}, yielding sub-percent level agreement with observational measurements. Turner et al. (2024), using DESI DR1, provided high-redshift optical depth constraints that enabled us to extend this optimization method to higher redshifts and generate corresponding mock spectra.
    
    In contrast, improvements in reproducing the \poned\ shape and amplitude result from fitting an analytic model to DESI EDR \cite{PD13, Karacayli_2020} and directly solving for the optimal baseline Gaussian correlation function at each redshift. This bypasses the limitations of generic parametric forms and allows our mocks to more accurately recover the scale dependence and amplitude of \poned\ over a wide range of $k$ and redshift.

    These synthetic mock spectra are well suited for a range of applications, particularly to support \poned\ measurements using fast Fourier transform (FFT) and likelihood-based estimators \cite{Ravoux2023, Karacayli_2024, Karacayli_2025}. By design, our mocks are uncontaminated, and each line-of-sight realization is independent and uncorrelated, making them ideal for controlled experiments that isolate the impact of observational effects. This enables systematic studies of factors such as spectral resolution, pixel masking, noise properties, and continuum estimation, all of which influence \poned\ analyses \cite[e.g.][]{Ravoux2023,Karacayli_2024}. Astrophysical contaminants -- such as high-column density systems (HCDs) and broad absorption lines (BALs) -- can be incorporated using tools like \textsc{quickquasars}\footnote{\url{https://github.com/desihub/desisim/blob/main/py/desisim/scripts/quickquasars.py}} from the \textsc{desisim}\footnote{\url{https://github.com/desihub/desisim}} package \cite{Herrera_Alcantar_2025}, which enables realistic forward modeling of contaminant populations in DESI quasar spectra.

    For DESI-like surveys, the dominant sources of systematic uncertainty in \poned\ measurements include damped \lya\ systems (DLAs), BALs, continuum estimation, spectrograph resolution, and noise modeling \cite{Karacayli_2025}. Among these, DLAs contribute the most significant bias on large scales due to their extended damping wings, which can enhance power at low $k$. False positives in the identification of DLAs can further exacerbate this bias. Although BAL contamination is typically an order of magnitude smaller than that of DLAs, it can still inflate \poned\ through correlated metal absorption features associated with quasar outflows -- features that are typically uncorrelated with the diffuse IGM. Errors in continuum fitting also introduce small biases ($\lesssim 1\%$) across all scales, with particularly strong effects at lower redshifts. 
    
    By leveraging uncontaminated mocks where the true signal is known, we can systematically evaluate these effects, improve contaminant identification, and quantify accurate error budgets for \poned\ analyses. More broadly, this framework supports robust cosmological parameter estimation, facilitates validation of analysis pipelines on survey data, and enables performance forecasting for future large-scale structure experiments by providing a controlled environment to evaluate and mitigate systematic effects. Furthermore, improving the agreement between the measured \poned\ of mocks and data creates a foundation for end-to-end analyses, such that emulator or simulation outputs for a given cosmology framework can be used as direct inputs for the generation of lognormal mocks. Since this lognormal method effectively replicates these inputs into transmission fields with accurate statistical properties, such an approach would enable controlled, survey-specific \poned\ predictions across a range of cosmologies, providing a direct bridge between theoretical models and observational inference.
    
    Our framework remains flexible and open to further refinement. An assumption of our model that merits improvement involves the treatment of the variance of the Gaussian field. As detailed in Appendix \ref{section: covariance matrix and variance}, our current method fixes $\sigma_F^2$ at a single value when we fit the mean flux, whereas the baryon field variance $\sigma_b^2$ varies with redshift and individual spectra. This may bias $\xi_G$ -- particularly at low redshifts ($z \lesssim 2.0$), where we observe an excess of power. A promising future direction would be to adopt a joint fit procedure for the mean flux and \poned, which would allow both to inform the model parameters simultaneously rather than in isolation.

\section{Conclusions} \label{Section: conclusion}
    We present and validate a fast and accurate lognormal mock generation framework for simulating one-dimensional \lya\ forest spectra, designed to support \poned\ analysis in large-scale surveys such as DESI. Our approach improves upon previous work by accurately reproducing the redshift evolution of the mean transmitted flux and the shape and amplitude of \poned\ across a wide range of scales and redshifts. These improvements are achieved by tuning key transformation parameters and directly fitting an underlying Gaussian correlation function, respectively.
    
    Our mocks recover the mean transmitted flux across the full DESI EDR redshift range ($2.0 \leq z \leq 3.8$) with a fractional RMS error of just 0.002. They also match the \poned\ shape and amplitude with an average RMS error of 0.02 over the DESI EDR $k$ range ($0.00075 \leq k \lesssim 0.035$ s km$^{-1}$). These gains are consistent with redshift, with residuals below 10\% for all redshifts at intermediate scales. The mocks maintain robust performance when extended to much broader scales ($10^{-5} \leq k \lesssim 0.1$ s km$^{-1}$), where the RMS remains just 0.1. 
    
    Unlike previous approaches, which perform well within restricted $k$-ranges tailored to specific measurements, our method generalizes effectively, and accurately replicates an input model or dataset across a broader range of scales. This flexibility makes it particularly well suited for future \lya\ forest studies that require high fidelity across a range of cosmological and instrumental configurations.
    
    Future work will focus on expanding the scientific utility of these mocks by incorporating astrophysical contaminants, continuum estimation uncertainties, and instrumental systematics. These enhancements will enable controlled systematics studies and broaden the applicability of the mocks to a variety of analyses. This framework would also improve masking corrections, following approaches similar to those proposed by Boera et al. (2019) \cite{Boera2019}, which leverages the comparison between contaminated and uncontaminated mock spectra to develop an empirical correction for various masking effects. This most notably includes \poned\ estimation with the FFT and optimal estimator pipelines \cite{Ravoux2023, Karacayli_2024} used for DESI \poned\ analysis, as well as the validation of forward-modeling approaches for quasar spectra \cite{Herrera_Alcantar_2025}. With continued refinement -- such as joint fitting of the mean flux and \poned\ -- this framework offers a robust, flexible foundation for generating large ensembles of synthetic spectra in support of precision cosmology with the \lya\ forest.

\acknowledgments

MH, NGK, and PM acknowledge support from the United States Department of Energy, Office of High Energy Physics under Award Number DE-SC0011726. This research used resources of the National Energy Research Scientific Computing Center (NERSC), a Department of Energy User Facility using NERSC award HEP-ERCAP0032748.

\appendix
\section{Software} \label{section: software}
    The code used to fit the mean flux evolution with redshift and the one-dimensional power for mock generation is publicly available\footnote{\url{https://github.com/m-herbold/P1D_Mocks}}. This code follows the method described in \S \ref{Section: fitting_power} and \S \ref{Section: fitting_mean_flux} and is designed to reproduce the mock generation process detailed in \S \ref{section: Mock_generation} by providing fitting parameters for the mean flux such as those in Table \ref{table.2} and a baseline model power spectrum $P_G(k,z)$ (from the Gaussian correlation function, $\xi_{G,fit}$) for the underlying one-dimensional Gaussian field. 

    We made extensive use of the \textsc{NumPy} \cite{harris2020array} and \textsc{SciPy} \cite{2020SciPy-NMeth} libraries for array manipulation and scientific computing. For nonlinear minimization, we used the \textsc{iminuit} package \cite{iminuit}, a Python interface to the Minuit2 C++ library. We used the \textsc{Astropy} library \cite{astropy:2013, astropy:2018, astropy:2022} for file handling and FITS I/O operations, and \textsc{pandas} \cite{mckinney-proc-scipy-2010, reback2020pandas} for tabular data manipulation and CSV handling. Plots and figures were generated using \textsc{Matplotlib} \cite{Hunter:2007}.

\section{Data Availability} \label{Section: Data Availability}
    Information and data associated with current and future DESI data releases, including the Early Data Release (EDR) and Data Release 1 (DR1) referenced in this work, are publicly available at \url{https://data.desi.lbl.gov/doc/releases/}. The \lya\ forest catalog for DESI EDR is available at \url{https://data.desi.lbl.gov/public/edr/vac/edr/lya/fuji/v0.3}. 

    The data points used in the figures are provided by \cite{Karacayli_2024}, and available as text files at \url{https://doi.org/10.5281/zenodo.8007370}.

\section{Covariance Matrix and Variance} \label{section: covariance matrix and variance}
    An underlying assumption of the covariance matrix in eq. \ref{eq.11} is that the variance $\sigma^2$ is equal to the zeroth velocity value of the Gaussian correlation function such that $\sigma^2 = \xi_G(v=0)$. This $v=0$ value for the Gaussian correlation function propagates to $v=0$ of the flux correlation function through the transformations outlined in \S \ref{section: Mock_generation}, such that $\xi_G(v=0) \mapsto \xi_F(v=0)$. Therefore, the assumption requires $\sigma^2$ to be greater than $\xi_G(v=0)$. Scenarios in which $\sigma^2 \sim \xi_G(0)$ will result in a cut-off in velocity, and thus an underestimate of the $v=0$ value in the correlation functions.
    
    Since the fitting method described in \S \ref{Section: fitting_mean_flux} utilizes $\sigma_F^2$ as a fitting parameter for the mean flux, and thus a constant value when fitting the one-dimensional power, it is not guaranteed that the relation $\sigma_F^2 > \xi_G(v=0)$ is always true. We choose a minimum velocity sufficiently greater than zero for fitting the one-dimensional power to enforce the condition that $\sigma_F^2 > \xi_G(v=0)$, while holding $\sigma_F^2$ constant. 
    
    We solve for $\xi_{G,fit}(v=0)$ independently of the mean flux and interpolate $\xi_{G,fit}$ to $v=0$ with piecewise polynomial interpolation\footnote{with scipy.interpolate.interpid \cite{2020SciPy-NMeth}}. This covariance matrix assumption is somewhat redshift dependent, and therefore requires redshift dependent limits on the minimum velocity value. We find that lower redshifts require higher minimum velocity value cut-offs, whereas higher redshifts tolerate smaller minimum velocity limits and therefore result in higher resolution fits on average.  
    
    We find that the difference in the preferred value of $\sigma^2$ by $\overline{F}(z)$ and $\xi_G(v=0)$ does not significantly affect the recovery of the target \poned. Therefore, it is sufficient to choose $v_{min}>0$, solve for $\xi_G(v=0)$ independently, and extrapolate to $v=0$. In a future work, we may investigate a joint fit of $\overline{F}(z)$ and $\xi_G(v=0)$ for a jointly preferred value of $\sigma^2$ at each redshift.

    \section{Convergence Analysis} \label{section: Convergence Analysis} 
    To determine the number of independent line-of-sight realizations ($N$) necessary for the validation tests described in \S \ref{section: Results}, we performed a convergence analysis on the \poned. Convergence was assessed using the root-mean-square (RMS) fractional error between our mock measurements and the DESI EDR target model, averaged over all redshifts ($2.0 \leq z \leq 3.8$).

    We find that the RMS fractional error asymptotically approaches a constant value for $N \approx 500$–$1000$ realizations, across all scales considered. To quantify this behavior, we define a rate of convergence as
    
    $$
    \text{Rate of Improvement per log step} = \frac{|S(N_{\text{old}}) - S(N_{\text{new}})|}{\log(N_{\text{new}})-\log(N_{\text{old}})},
    $$
    where $S(N)$ denotes the RMS fractional error computed using $N$ realizations. This finite-difference measure captures how rapidly the statistic improves as the sample size increases. As this rate approaches zero, additional realizations provide negligible improvement, indicating convergence.

    Applying this criterion, we find that the convergence rate of the \poned\ RMS fractional error becomes effectively zero at $N = 1000$.

\bibliographystyle{JHEP}
\bibliography{biblio}

\end{document}